\documentclass{article}
\usepackage{amsmath,graphicx}
\usepackage[preprint]{spconf}
\toappear{\begin{tabular}{l}This work has been submitted to the IEEE ICASSP 2023 for possible publication. \\
	Copyright may be transferred without notice, after which this version may no longer be accessible.\end{tabular}}
\usepackage{bm}
\usepackage{amssymb}
\usepackage[skip=5pt]{caption}

\DeclareMathOperator{\diag}{diag}
\usepackage{amsmath}
\newcommand\norm[1]{\lVert#1\rVert}
\usepackage{tikz}
\usetikzlibrary{calc, angles, quotes}
\usetikzlibrary{calc, shapes.geometric}
\usepackage{pgfplots}
\usetikzlibrary{backgrounds}
\usepackage[normalem]{ulem}

\usepackage[acronym,shortcuts]{glossaries}

\newacronym{LMMSE}{LMMSE}{linear minimum mean squared error}
\newacronym{MSE}{MSE}{mean squared error}
\newacronym{MIMO}{MIMO}{multiple-input multiple-output}
\newacronym{SISO}{SISO}{single-input single-output}
\newacronym{DL}{DL}{deep learning}
\newacronym{OFDM}{OFDM}{orthogonal frequency division multiplexing}
\newacronym{CSI}{CSI}{channel state information}
\newacronym{ULA}{ULA}{uniform linear array}
\newacronym{DFT}{DFT}{discrete fourier transform}
\newacronym{BS}{BS}{base station}
\newacronym{MT}{MT}{mobile terminal}
\newacronym{VAE}{VAE}{variational autoencoder}
\newacronym{GMM}{GMM}{Gaussian mixture model}
\newacronym{SIMO}{SIMO}{single-input multiple-output}
\newacronym{ML}{ML}{machine learning}
\newacronym{DoA}{DoA}{direction of arrival}
\newacronym{KL}{KL}{Kullback-Leibler}
\newacronym{ELBO}{ELBO}{evidence-lower bound}
\newacronym{i.i.d.}{i.i.d.}{independent and identically distributed}
\newacronym{MMSE}{MMSE}{minimum mean squared error}
\newacronym{DVAE}{DVAE}{Dynamical VAE}
\newacronym{kMMVAE}{$k$-MMVAE}{$k$-MemoryMarkovVAE}
\newacronym{FC}{FC}{fully connected}
\newacronym{CNN}{CNN}{convolutional neural network}
\newacronym{LS}{LS}{least squares}
\newacronym{NMSE}{NMSE}{normalized mean squared error}
\newacronym{SNR}{SNR}{signal-to-noise ratio}
\newacronym{CE}{CE}{channel estimation}

\title{VARIATIONAL INFERENCE AIDED ESTIMATION OF TIME VARYING CHANNELS}
\name{Benedikt B\"ock, Michael Baur, Valentina Rizzello, Wolfgang Utschick}
\address{School of Computation, Information and Technology, Technical University of Munich, Germany\\
	Email: \{benedikt.boeck, mi.baur, valentina.rizzello, utschick\}@tum.de}
\begin{document}
	\ninept
	\maketitle
	\begin{abstract}
		One way to improve the estimation of time varying channels is to incorporate knowledge of previous observations. In this context, \acp{DVAE} build a promising \ac{DL} framework which is well suited to learn the distribution of time series data. We introduce a new \ac{DVAE} architecture, called \ac{kMMVAE}, whose sparsity can be controlled by an additional memory parameter. Following the approach in~\cite{VAEPaper} we derive a \ac{kMMVAE} aided channel estimator which takes temporal correlations of successive observations into account. 
		The results are evaluated on simulated channels by QuaDRiGa and show that the \ac{kMMVAE} aided channel estimator clearly outperforms other \ac{ML} aided estimators which are either memoryless or naively extended to time varying  channels  without major adaptions.
	\end{abstract}
	\section{Introduction}
	\label{sec:intro}
	Recently, \ac{DL} aided \ac{CE} has demonstrated great performance in a variety of wireless communication systems \cite{VAEPaper}\nocite{SISOEstimation1}\nocite{SISOEstimation2}\nocite{MIMOEstimation}-\cite{GMMPaper}.
	Despite following the general \ac{DL} paradigm of utilizing data to learn functional dependencies, the core idea of the introduced methods can be divided into two different categories - end-to-end learning and model based learning. In the former the input-output relation of interest is treated as a black box and learned by a neural network while in the latter, isolated parts of a classically modelled relation are replaced by neural networks \cite{ModelBasedDeepLearning}. Generative models like \acp{VAE} and \acp{GMM} represent promising \ac{ML} tools which are well suited for model based learning and have shown good results for channel estimation in massive \ac{MIMO} systems \cite{VAEPaper,GMMPaper}. In these setups, a learnable function maps a channel observation from a fixed environment to the statistical parameters of a so called latent random vector which can be associated with scenario specific characteristics of the corresponding radio propagation environment. 
	The channel can be assumed to be conditionally Gaussian with respect to these characteristics. By utilizing the law of total expectation a parameterized \ac{LMMSE} estimator can be derived, which is approximately \ac{MSE} optimal although the channel can follow any arbitrary distribution. These approaches are solely based on instantaneous \ac{CSI} and do not take any temporal correlation into account. However, the temporal evolution of the channel is a scenario specific characteristic by itself and further performance gains are expected by incorporating it into a \ac{DL} aided model for channel estimation.
	
	Our contribution in this work is to extend the \ac{VAE} framework in \cite{VAEPaper} to estimate the channel based on correlated channel observations along a user's trajectory. In this context, we introduce a new \ac{DL} architecture, called \ac{kMMVAE}, which is motivated by the framework of \acp{DVAE} such as KalmanVAEs and Deep Kalman Filters \cite{KalmanVAE,DeepKalmanFilter}. \acp{DVAE} model the latent random vector as a Markov chain and are therefore well suited to represent the temporal correlation in the input data \cite{DVAEsReview}. In contrast to other architectures, our proposed model incorporates an additional memory parameter controlling the sparsity of the corresponding probabilistic graph. Moreover, the \ac{VAE} aided estimator in \cite{VAEPaper} as well as the \ac{kMMVAE} are both generalized to take noisy observations with arbitrary \acp{SNR} as input. We show that already for very short trajectories the extracted additional information about the temporal evolution leads to significant better estimates compared to the memoryless case. Additionally, we compare our proposed model with a model-agnostic approach, where we process time correlated data from trajectories via an ordinary \ac{VAE} leading to a noticeable performance decrease.
	
	\section{System and Channel Model}
	\label{sec:system_model}
	The considered system is a \ac{SIMO} setup, in which the \ac{BS} is equipped with $R$ antennas and receives uplink training signals from a single antenna \ac{MT}. We concentrate on $I$ consecutive pilot symbols which are temporally spaced by a time interval $T$. After decorrelating the pilots, the resulting received signal $\bm{y}_i$ at time $iT$ can be expressed as
	\begin{equation}
		\label{aquisition_samples}
		\bm{y}_i = \bm{h}_i + \bm{n}_i \in \mathbb{C}^R, i=1,\ldots,I.
	\end{equation}
	The channel vector $\bm{h}_i$ is perturbed by additive white Gaussian noise  $\bm{n}_i$ at time $iT$, i.e., $\bm{n}_i \sim \mathcal{N}_{\mathbb{C}}(\mathbf{0},\sigma_n^2\mathbf{I})$ and $\mathbb{E}[\bm{n}_i\bm{n}_{\tilde{i}}^\text{H}] = \mathbf{0}$ for $i \neq \tilde{i}$. We assume the channel vectors of different snapshots to be correlated and the \ac{BS} antennas to form a \ac{ULA} with half-wavelength spacing. It is known that the resulting channel covariance matrix at any time $iT$ is Toeplitz structured and can be approximated by a circulant matrix $\bm{C}_i$ for a large number of antennas \cite{CirculantApprox}. Consequently, we can diagonalize $\bm{C}_i$ by utilizing the fact that the eigenvectors of any circulant matrix coincide with the columns of the \ac{DFT}-matrix, i.e.,
	\begin{equation}
		\bm{C}_i = \bm{F}^{\text{H}}\diag(\bm{c}_i)\bm{F},
	\end{equation}
	where $\bm{F}$ stands for the R$\times$R \ac{DFT}-matrix.
	
	The channel realizations for training and evaluating the proposed model are generated by the geometry-based stochastic channel modeling tool QuaDRiGa \cite{QuadrigaPaper,QuadrigaDoc}. There, the time dependent channel for a fixed center frequency $f_c$ is modeled as a superposition of $L$ distinct propagation paths, i.e.,
	\begin{equation}
		\bm{h}_i = \sum_{l=0}^{L-1} \bm{g}_l(iT) \exp(-\text{j}2 \pi f_c\tau_l(iT)),
	\end{equation}
	where $\tau_l(iT)$ represents the path delay and $\bm{g}_l(iT)$ contains information about path gain, \ac{DoA}, polarization effects and subpath characteristics for path $l$ at time $iT$. The initial values of all these parameters are drawn from a scenario specific distribution and are then coherently updated for all snapshots along the user's trajectory. This procedure results in an environment specific time evolution of the channel. 
	
	\section{Variational Inference Aided Channel Estimation}
	\subsection{VAE Preliminaries}
	\label{subs:VAE_preliminaries}
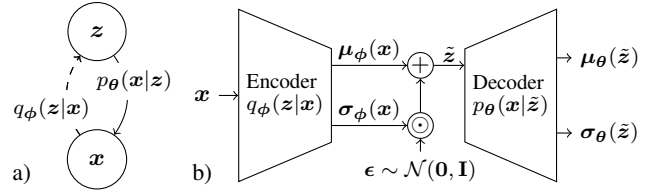
\begin{figure}[t]
		\begin{center}
				\begin{tikzpicture}
		\def\Blocksize{1.4cm}
		\def\Blockwidth{1cm}
		\def\Blockheight{2.3cm}
		\def\Blockdistance{3.cm}
		\def\Blockangle{75}
		
		\node (input) at (2, 0) {$\bm{x}$};
		
		\node[trapezium, draw, align=center, trapezium stretches = true, minimum height=\Blockwidth, minimum width=\Blockheight, align=center, trapezium angle=\Blockangle, rotate=-90] (NN1) at($ (input) + (1.1,0) $) {\rotatebox{90}{\footnotesize Encoder}\vspace{5mm}{ \rotatebox{90}{\footnotesize $q_{\bm{\phi}}(\bm{z}|\bm{x})$}}};
		
		\node[trapezium, draw, align=center, trapezium stretches = true, minimum height=\Blockwidth, minimum width=\Blockheight, align=center, trapezium angle=\Blockangle, rotate=90] (NN2) at($ (NN1) + (\Blockdistance,0) $) {{\rotatebox{-90}{\footnotesize $p_{\bm{\theta}}(\bm{x}|\tilde{\bm{z}})$}} \vspace{10mm}\normalsize{\rotatebox{-90}{\footnotesize Decoder}}};

		\node[circle,draw,inner sep = 0.01em] (add) at (2 + 2.9,0.4){$+$};
		\node[circle,draw,inner sep = 0.01em] (multiply) at (2 + 2.9,-0.4){$\odot$};
		
		\node (epsilon) at (2 + 2.9,-1){\footnotesize $\bm{\epsilon} \sim \mathcal{N}(\bm{0},\mathbf{I})$};
		
		\node (outputmu) at ($(NN2.south) + (0.7,0.5)$) {\footnotesize $\bm{\mu}_{\bm{\theta}}(\tilde{\bm{z}})$};
		\node (outputcov) at ($(NN2.south) + (0.7,-0.5)$) {\footnotesize $\bm{\sigma}_{\bm{\theta}}(\tilde{\bm{z}})$};

		\draw[->] ($ (NN1.north) + (0,0.4)$) --node[midway,above = -0.2em]{\footnotesize $\bm{\mu}_{\bm{\phi}}(\bm{x})$} (add);
		\draw[->] ($ (NN1.north) + (0,-0.4)$) --node[midway,above = -0.2em]{\footnotesize $\bm{\sigma}_{\bm{\phi}}(\bm{x})$} (multiply);
		\draw[->] ($(add.east)$) --node[midway,above = -0.2em]{\footnotesize $\tilde{\bm{z}}$} ($(NN2.north) + (0,0.4)$);
		\draw[->] ($(epsilon.north)$) -- ($(multiply.south)$);
		\draw[->] ($(multiply.north)$) -- ($(add.south)$);
		\draw[->] ($(NN2.south) + (0,0.5)$) -- ($(outputmu.west)$);
		\draw[->] ($(NN2.south) + (0,-0.5)$) -- ($(outputcov.west)$);
		\draw[->] ($(input.east)$) -- ($(NN1.south)$);
		
		\node[circle,draw,inner sep = 0.6em] (x) at (0.6,-0.85) {$\bm{x}$};	
		\node[circle,draw,inner sep = 0.6em] (z) at (0.6,0.85) {$\bm{z}$};	
	
		\node (a) at ($(x.south) + (-1,0.2)$) {a)};
		\node (b) at ($(x.south) + (1.4,0.2)$) {b)};
		
		\draw[->] (z) to[out=-55,in=55] (x);
		\draw[dashed,->] (x) to[out=125,in=-125] (z);
		\draw[dashed,->] (x) to[out=125,in=-125] (z);
		
		\node[fill=white] (q) at (0,-0.19) {\footnotesize $q_{\bm{\phi}}(\bm{z}|\bm{x})$};
		\node[fill=white] (p) at (1.1,0.19) {\footnotesize $p_{\bm{\theta}}(\bm{x}|\bm{z})$};
	\end{tikzpicture}
		\end{center}
		\caption{Probabilistic graph (a) and architecture (b) of the standard \ac{VAE} framework.}
		\label{fig:ordinaryVAE}
		\vspace{-9pt}
	\end{figure}

	Generative models and \acp{VAE} in particular aim to learn a distribution $p(\bm{x})$ based on a dataset $\mathcal{X} = \{\bm{x}^{(n)}\}_{n=1}^{N}$ by applying likelihood estimation to a parameterized statistical model \cite{introVAE}. In order to increase the expressiveness of this model, \acp{VAE} introduce a low dimensional non observable (i.e. latent) random vector $\bm{z}$, such that for each $\bm{x}^{(n)}$ there exists a corresponding realization $\bm{z}^{(n)}$ of $\bm{z}$ which summarizes the key features of $\bm{x}^{(n)}$ \cite{originalVAE}. In the standard \ac{VAE} framework, the distribution of $\bm{z}$, called prior distribution, is fixed (e.g. $\mathcal{N}(\mathbf{0},\mathbf{I})$) and the conditional distribution $p_{\bm{\bm{\theta}}}(\bm{x}|\bm{z})$ is parameterized by a learnable parameter $\bm{\theta}$ representing e.g. the weights of a neural network. A drawback of introducing $\bm{z}$ is that the resulting marginalized likelihood $p_{\bm{\theta}}(\bm{x})$ is not computable due to the intractability of $p_{\bm{\theta}}(\bm{z}|\bm{x})$. To overcome this issue, $p_{\bm{\theta}}(\bm{z}|\bm{x})$ is approximately inferred by minimizing the \ac{KL} divergence between $p_{\bm{\theta}}(\bm{z}|\bm{x})$ and a tractable distribution $q_{\bm{\phi}}(\bm{z}|\bm{x})$. The parameter $\bm{\phi}$ is learnable and stands for e.g. the weights of a neural network. By subtracting this \ac{KL} divergence from the log likelihood, the resulting expression is a tractable lower bound on the log likelihood $\log p_{\bm{\theta}}(\bm{x})$ called \ac{ELBO} $\mathcal{L}(\bm{\theta},\bm{\phi})$. Assuming that the samples in $\mathcal{X}$ are \ac{i.i.d.}, an equivalent and more practical version of the \ac{ELBO} is given by
	\begin{equation}
		\label{eq:ELBO}
		\mathcal{L}(\bm{\theta},\bm{\phi}) = \sum_{n=1}^N \mathbb{E}_{q_{\bm{\phi}}(\bm{z}|\bm{x}^{(n)})}\left[\log p_{\bm{\theta}}(\bm{x}^{(n)}|\bm{z}) - \log \frac{q_{\bm{\phi}}(\bm{z}|\bm{x}^{(n)})}{p(\bm{z})}\right].
	\end{equation}
	Usually, the expectation is approximated by a Monte-Carlo estimation based on a single sample drawn from $q_{\bm{\phi}}(\bm{z}|\bm{x}^{(n)})$. The relation between $\bm{x}$ and $\bm{z}$ is represented as probabilistic graph in Fig.~\ref{fig:ordinaryVAE}~a), in which the parameterized statistical model $p_{\bm{\bm{\theta}}}(\bm{x}|\bm{z})$ is illustrated by the solid arrow and can be distinguished from the approximate inference distribution $q_{\bm{\phi}}(\bm{z}|\bm{x})$ displayed as a dashed arrow. In standard \acp{VAE}, $q_{\bm{\phi}}(\bm{z}|\bm{x})$, as well as $p_{\bm{\theta}}(\bm{x}|\bm{z})$, are modeled as conditionally Gaussian distributions with means $\bm{\mu}_{\bm{\theta}}(\bm{z})$ and $\bm{\mu}_{\bm{\phi}}(\bm{x})$, and diagonal covariance matrices $\diag(\bm{\sigma}_{\bm{\theta}}(\bm{z})^2)$ and $\diag(\bm{\sigma}_{\bm{\phi}}(\bm{x})^2)$, respectively. 
	
	The corresponding architecture is illustrated in Fig.~\ref{fig:ordinaryVAE}~b). The parameterization $\bm{\phi}$ is realized by a neural network, called encoder, which takes a sample $\bm{x}$ as input and outputs $\bm{\mu}_{\bm{\phi}}(\bm{x})$, as well as $ \bm{\sigma}_{\bm{\phi}}(\bm{x})$. Subsequently, a single sample $\tilde{\bm{z}}$ is drawn from $q_{\bm{\phi}}(\bm{z}|\bm{x})$ by computing it via $\tilde{\bm{z}} = \bm{\mu}_{\bm{\phi}}(\bm{x}) + \bm{\sigma}_{\bm{\phi}}(\bm{x}) \odot \bm{\epsilon}$  where $\bm{\epsilon} \sim \mathcal{N}(\mathbf{0},\mathbf{I})$. This is known as the reparameterization trick. Eventually, a second neural network, called decoder, takes $\tilde{\bm{z}}$ as input and outputs $\bm{\mu}_{\bm{\theta}}(\tilde{\bm{z}})$ as well as $\bm{\sigma}_{\bm{\theta}}(\tilde{\bm{z}})$. Based on these parameters, the \ac{ELBO} can be evaluated and maximized by a gradient based optimization algorithm. 
	\subsection{VAE Based \ac{CE} with Instantenous CSI}
	\label{VAE_CE_instantenous_CSI}
	There are several ways how to utilize standard \acp{VAE} for channel estimation based on instantaneous \ac{CSI}. In this section, we give a brief overview of the key concepts by focusing on one particular way and refer to \cite{VAEPaper} for a more detailed explanation. The data set $\mathcal{X}$ consists of \ac{i.i.d.} channel realizations $\{\bm{h}^{(n)}\}_{n=1}^N$ at single time instances. In this case, the approximate inference distribution $q_{\bm{\phi}}(\bm{z}|\cdot)$ is conditioned on noisy observations $\bm{y}$ of the samples in $\mathcal{X}$ according to \eqref{aquisition_samples}. Consequently, perfect \ac{CSI} in form of noiseless channel realizations is only required during the training phase. The law of total expectation
	\begin{equation}
		\label{eq:law_total_expectation}
		\mathbb{E}_{p(\bm{h}|\bm{y})}[\bm{h}|\bm{y}] = \mathbb{E}_{p(\bm{z}|\bm{y})}[\mathbb{E}_{p(\bm{h}|\bm{y},\bm{z})}[\bm{h}|\bm{y},\bm{z}]]
	\end{equation}
	is used to reformulate the \ac{MMSE} estimator as an expected \ac{LMMSE} estimator
	\begin{equation}
		\label{eq:LMMSE_MMSE}
		\mathbb{E}_{p(\bm{z}|\bm{y})}[\bm{\mu}_{\bm{h}|\bm{z}} + \bm{C}_{\bm{h}|\bm{z}} (\bm{C}_{\bm{h}|\bm{z}} + \sigma_n^2 \mathbf{I})^{-1} (\bm{y} - \bm{\mu}_{\bm{h}|\bm{z}})].
	\end{equation}
	The parameters $\bm{\mu}_{\bm{h}|\bm{z}}$ and $\bm{C}_{\bm{h}|\bm{z}}$ are mean and covariance matrix of the channel $\bm{h}$ conditioned on the latent random vector $\bm{z}$.
	As explained in Section~\ref{sec:system_model}, $\bm{C}_{\bm{h}|\bm{z}}$ can be assumed to be diagonal by preprocessing the noiseless as well as the noisy channel realizations by a DFT transformation.  The estimator in \eqref{eq:LMMSE_MMSE} can therefore be approximated by an \ac{ML} aided \ac{LMMSE} estimator for which the expectation operation can be dropped by replacing $\bm{\mu}_{\bm{h}|\bm{z}}$ and $\bm{C}_{\bm{h}|\bm{z}}$ with the \ac{VAE} learned parameters $\bm{\mu}_{\bm{\theta}}(\bm{\mu}_{\bm{\phi}}(\bm{y}))$ and $\diag(\bm{\sigma}_{\bm{\theta}}(\bm{\mu}_{\bm{\phi}}(\bm{y}))^2)$.
	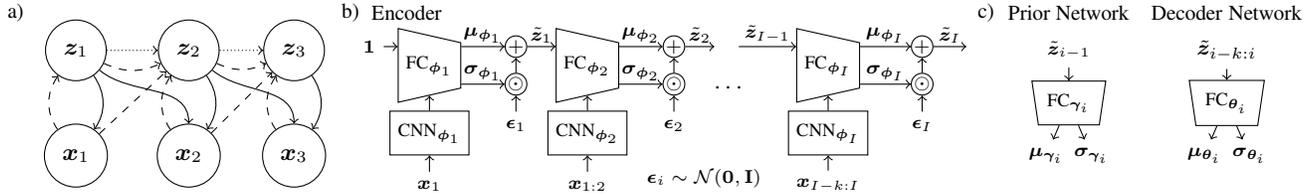
\begin{figure*}[t]%
		\begin{center}
				\begin{tikzpicture}
		\def\Blocksize{0.9cm}
		\def\Blockwidth{0.6cm}
		\def\Blockheight{1cm}
		\def\Blockdistance{3.cm}
		\def\Blockangle{80}
		
		\node[circle,draw,inner sep = 0.5em] (x1) at (-0.2,-0.6+4) {$\bm{x}_1$};	
		\node[circle,draw,inner sep = 0.5em] (z1) at (-0.2,0.8+4) {$\bm{z}_1$};	
		\node[circle,draw,inner sep = 0.5em] (x2) at (1.3,-0.6+4) {$\bm{x}_2$};	
		\node[circle,draw,inner sep = 0.5em] (z2) at (1.3,0.8+4) {$\bm{z}_2$};	
		\node[circle,draw,inner sep = 0.5em] (x3) at (2.7,-0.6+4) {$\bm{x}_3$};	
		\node[circle,draw,inner sep = 0.5em] (z3) at (2.7,0.8+4) {$\bm{z}_3$};	
	
		\node (enctotal) at (4.2,5.3) {\footnotesize Encoder};
		
		\node[draw, align=center, trapezium stretches = true, minimum height=\Blockwidth-0.5, minimum width=\Blockheight-0.5, align=center, trapezium angle=\Blockangle, rotate=-180] (ENCODER0) at ($(4.5,3.7)$) {\rotatebox{180}{\scriptsize $\text{CNN}_{\bm{\phi}_1}$}};
		\node[trapezium, draw, align=center, trapezium stretches = true, minimum height=\Blockwidth-0.5, minimum width=\Blockheight-0.5, align=center, trapezium angle=\Blockangle, rotate=-90] (ENCODER20) at ($(4.5,4.6)$) {\rotatebox{90}{\scriptsize $\text{FC}_{\bm{\phi}_1}$}};
		\node (EncIn10) at (4.5,3.4-0.4) {\scriptsize $ \bm{x}_{1}$};
		\node (EncIn20) at (3.7,5.25-0.4) {\scriptsize $\bm{1}$};
		\draw[->] (EncIn10) to (ENCODER0);
		\draw[->] (EncIn20) to ($(ENCODER20.south) + (0,0.25)$);
		\node[circle,draw,inner sep = 0.005em] (add0) at (5.65,5.25-0.4){\scriptsize $+$};
		\node[circle,draw,inner sep = 0.005em] (multiply0) at (5.65,4.75-0.4){\scriptsize $\odot$};
		\draw[->] ($(ENCODER20.north) + (0,0.25)$) -- node[midway,above = -0.2em]{\scriptsize $\bm{\mu}_{\bm{\phi}_1}$} (add0.west);
		\draw[->] (ENCODER0.south) to (ENCODER20.east);
		\draw[->] ($(ENCODER20.north) - (0,0.25)$) -- node[midway,above = -0.2em]{\scriptsize $\bm{\sigma}_{\bm{\phi}_1}$} (multiply0.west);
		
		\node (epsilon0) at (5.65,4.2-0.4){\scriptsize $\bm{\epsilon}_1$};
		
		\node[draw, align=center, trapezium stretches = true, minimum height=\Blockwidth-0.5, minimum width=\Blockheight-0.5, align=center, trapezium angle=\Blockangle, rotate=-180] (ENCODER1) at ($(6.6,4.1-0.4)$) {\rotatebox{180}{\scriptsize $\text{CNN}_{\bm{\phi}_2}$}};
		\node[trapezium, draw, align=center, trapezium stretches = true, minimum height=\Blockwidth-0.5, minimum width=\Blockheight-0.5, align=center, trapezium angle=\Blockangle, rotate=-90] (ENCODER21) at ($(6.6,5-0.4)$) {\rotatebox{90}{\scriptsize $\text{FC}_{\bm{\phi}_2}$}};
		\node (EncIn11) at (6.6,3.4-0.4) {\scriptsize $ \bm{x}_{1:2}$};
		\draw[->] (EncIn11) to (ENCODER1);
		\node[circle,draw,inner sep = 0.005em] (add1) at (7.75,5.25-0.4){\scriptsize $+$};
		\node[circle,draw,inner sep = 0.005em] (multiply1) at (7.75,4.75-0.4){\scriptsize $\odot$};
		\draw[->] ($(ENCODER21.north) + (0,0.25)$) -- node[midway,above = -0.2em]{\scriptsize $\bm{\mu}_{\bm{\phi}_2}$} (add1.west);
		\draw[->] (ENCODER1.south) to (ENCODER21.east);
		\draw[->] ($(ENCODER21.north) - (0,0.25)$) -- node[midway,above = -0.2em]{\scriptsize $\bm{\sigma}_{\bm{\phi}_2}$} (multiply1.west);
		
		\node (epsilon1) at (7.75,4.2-0.4){\scriptsize $\bm{\epsilon}_2$};
		
		\draw[->] (multiply0.north) to (add0.south);
		\draw[->] (multiply1.north) to (add1.south);
		\draw[->] (epsilon0) to (multiply0.south);
		\draw[->] (epsilon1) to (multiply1.south);
		\draw[->] (add0.east) -- node[midway,above = -0.2em]{\scriptsize $\tilde{\bm{z}}_1$} ($(ENCODER21.south) + (0,0.25)$);
		
		\node(nichts) at (8.4,5.25-0.4){};
		\draw[->] (add1.east) -- node[midway,above = -0.2em]{\scriptsize $\tilde{\bm{z}}_2$} (nichts);

		\node[draw, align=center, trapezium stretches = true, minimum height=\Blockwidth-0.5, minimum width=\Blockheight-0.5, align=center, trapezium angle=\Blockangle, rotate=-180] (ENCODER3) at ($(9.8,4.1-0.4)$) {\rotatebox{180}{\scriptsize $\text{CNN}_{\bm{\phi}_I}$}};
		\node[trapezium, draw, align=center, trapezium stretches = true, minimum height=\Blockwidth-0.6, minimum width=\Blockheight-0.6, align=center, trapezium angle=\Blockangle, rotate=-90] (ENCODER23) at ($(9.8,5-0.4)$) {\rotatebox{90}{\scriptsize $\text{FC}_{\bm{\phi}_I}$}};
		\node (EncIn13) at (9.8,3.4-0.4) {\scriptsize $ \bm{x}_{I-k:I}$};
		\draw[->] (EncIn13) to (ENCODER3);
		\node[circle,draw,inner sep = 0.005em] (add3) at (11.05,5.25-0.4){\scriptsize $+$};
		\node[circle,draw,inner sep = 0.005em] (multiply3) at (11.05,4.75-0.4){\scriptsize $\odot$};
		\draw[->] ($(ENCODER23.north) + (0,0.25)$) -- node[midway,above = -0.2em]{\scriptsize $\bm{\mu}_{\bm{\phi}_I}$} (add3.west);
		\draw[->] (ENCODER3.south) to (ENCODER23.east);
		\draw[->] ($(ENCODER23.north) - (0,0.25)$) -- node[midway,above = -0.2em]{\scriptsize $\bm{\sigma}_{\bm{\phi}_I}$} (multiply3.west);
		
		\node (epsilon3) at (11.05,4.2-0.4){\scriptsize $\bm{\epsilon}_I$};
		\node(nichts2) at (11.75,5.25-0.4){};
		\draw[->] (add3.east) -- node[midway,above = -0.2em]{\scriptsize $\tilde{\bm{z}}_I$} (nichts2);
		\draw[->] (multiply3.north) to (add3.south);
		\draw[->] (epsilon3) to (multiply3.south);
		
		\node(nichtsIn) at (8.5,5.25-0.4){};
		\draw[->] (nichtsIn.east) -- node[midway,above = -0.2em]{\scriptsize $\tilde{\bm{z}}_{I-1}$} ($(ENCODER23.south) + (0,0.25)$);

		\node(Punkt) at (8.5,4.7-0.4){$\cdots$};
		
		\node(eps) at (8.15,3.1){\scriptsize $\bm{\epsilon}_i \sim \mathcal{N}(\bm{0},\mathbf{I})$};
		
		\draw[->] (z1) to[out=-55,in=55] (x1);
		\draw[dashed,->] (x1) to[out=125,in=-125] (z1);
		
		\draw[->] (z2) to[out=-55,in=55] (x2);
		\draw[dashed,->] (x2) to[out=125,in=-125] (z2);;
		
		\draw[->] (z3) to[out=-55,in=55] (x3);
		\draw[dashed,->] (x3) to[out=125,in=-125] (z3);

		\draw[densely dotted,->] (z1) to (z2);
		\draw[densely dotted,->] (z2) to (z3);

		\draw[dashed,->] (z1) to[out=-25,in=-155] (z2);
		\draw[dashed,->] (z2) to[out=-25,in=-155] (z3);

		\draw[dashed,->] (x1) to (z2);
		\draw[dashed,->] (x2) to (z3);
		
		\draw[->] (z2) to[out=-40,in=90] (x3);
		\draw[->] (z1) to[out=-40,in=90] (x2);
		
		\node (priorNet) at (13,5.3) {\footnotesize Prior Network};
		\node (PriorIn) at (13,4.8) {\scriptsize $ \tilde{\bm{z}}_{i-1}$};
		\node[trapezium, draw, align=center, trapezium stretches = true, minimum height=\Blockwidth-0.5, minimum width=\Blockheight-0.5, align=center, trapezium angle=\Blockangle, rotate=-180] (PRIOR) at ($(13,4.1)$) {\rotatebox{180}{\scriptsize $\text{FC}_{\bm{\gamma}_i}$}};
		\node (muPrior) at (12.7,3.4) {\scriptsize $\bm{\mu}_{\bm{\gamma}_i}$};
		\node (sigPrior) at (13.3,3.4) {\scriptsize $\bm{\sigma}_{\bm{\gamma}_i}$};
		\draw[->] (PRIOR) to (muPrior);
		\draw[->] (PRIOR) to (sigPrior);
		\draw[->] (PriorIn) to (PRIOR);

		\node (decNet) at (15.1,5.3) {\footnotesize Decoder Network};
		\node (DecIn) at (15.1,4.8) {\footnotesize $ \tilde{\bm{z}}_{i-k:i}$};
		\node[trapezium, draw, align=center, trapezium stretches = true, minimum height=\Blockwidth, minimum width=\Blockheight, align=center, trapezium angle=\Blockangle, rotate=-180] (DECODER) at ($(15.1,4.1)$) {\rotatebox{180}{\scriptsize $\text{FC}_{\bm{\theta}_i}$}};
		\node (muDec) at (14.8,3.4) {\scriptsize $\bm{\mu}_{\bm{\theta}_i}$};
		\node (sigDec) at (15.4,3.4) {\scriptsize $\bm{\sigma}_{\bm{\theta}_i}$};
		\draw[->] (DECODER) to (muDec);
		\draw[->] (DECODER) to (sigDec);
		\draw[->] (DecIn) to (DECODER);
		
		\node (a) at (-1,5.3) {\footnotesize a)};
		\node (b) at (3.45,5.3) {\footnotesize b)};
		\node (c) at (11.9,5.3) {\footnotesize c)};
		
	\end{tikzpicture}
		\end{center}
		\caption{Probabilistic graph (a), encoder architecture (b), and decoder and prior network architecture (c) of the \ac{kMMVAE}.}
		\vspace{-9pt}
		\label{fig:DVAE}
	\end{figure*}
	\subsection{Dynamical VAE Preliminaries}
	The standard \ac{VAE} enforces the entries of $\bm{z}$ as well as $\bm{x}|\bm{z}$ to be mutually independent by modeling the corresponding covariance matrices to be diagonal. However, there are applications for which this assumption does not reflect the actual structure of the given data. One example are stochastic processes, where a single sample $\bm{x}$ consists of $I$ successively sampled and potentially correlated values $\bm{x}_i$, i.e., $\bm{x} = [\bm{x}_1^\mathrm{T},\ldots,\bm{x}_I^\mathrm{T}]^\mathrm{T}$. \acp{DVAE} form an extended \ac{VAE} framework and consider temporal correlations within a single sample $\bm{x}$ by introducing dependencies between entries of the latent random vector $\bm{z}$~\cite{DVAEsReview}. A common choice is to model $\bm{z}$ as a Markov chain with parameterized transition probabilities, i.e.,
	\begin{equation}
		\label{eq:DVAE_prior}
		p_{\bm{\gamma}}(\bm{z}) = \prod_{i=1}^I p_{\bm{\gamma}_i}(\bm{z}_i|\bm{z}_{i-1}),
	\end{equation}
	where $\bm{z} = [\bm{z}_1^\mathrm{T},\ldots,\bm{z}_I^\mathrm{T}]^\mathrm{T}$ and $\bm{z}_0 = \emptyset$. The parameter $\bm{\gamma} = [\bm{\gamma}_1^\mathrm{T},\ldots,\bm{\gamma}_I^\mathrm{T}]^\mathrm{T}$ is learnable and can be implemented in different ways. One example is the KalmanVAE, in which $\bm{\gamma}_i$ parameterizes a linear state space model \cite{KalmanVAE}. In contrast, Deep Kalman Filters realize $\bm{\gamma}_i$ by a neural network which outputs the mean and covariance matrix of $\bm{z}_i|\bm{z}_{i-1}$ \cite{DeepKalmanFilter}. Besides the distribution of $\bm{z}$, $q_{\bm{\phi}}(\bm{z}|\bm{x})$ and $p_{\bm{\theta}}(\bm{z}|\bm{x})$ are also tailored to the given data structure and can be decomposed similarly to $p_{\bm{\gamma}}(\bm{z})$ in \eqref{eq:DVAE_prior}. By modeling the latent space as a Markov chain according to \eqref{eq:DVAE_prior}, $\bm{z}_i$ depends on the realization of $\bm{z}_{i-1}$ which is commonly incorporated in the decomposition of $q_{\bm{\phi}}(\bm{z}|\bm{x})$. Additionally, the \ac{ELBO} is adapted and an iterative sampling procedure is introduced keeping the computation of the \ac{ELBO} efficient. This is explained in more detail in Section \ref{sec:HiddenMarkovVAE} based on our proposed model.
	
	\subsection{$k$-MemoryMarkovVAE}
	\label{sec:HiddenMarkovVAE}
	In this work, we propose an adapted DVAE architecture, called $k$-MemoryMarkovVAE, which to the best of our knowledge, has not been introduced in other publications. Just like KalmanVAEs and Deep Kalman Filters, \acp{kMMVAE} model the latent space as a Markov chain according to \eqref{eq:DVAE_prior}. In contrast to other \acp{DVAE} however, we introduce an adjustable hyperparameter $k$ standing for the memory of $q_{\bm{\phi}}(\bm{z}|\bm{x})$ and $p_{\bm{\theta}}(\bm{z}|\bm{x})$. More precisely, these distributions are decomposed as
	\begin{align}
		\label{eq:HMVAE_encoder}
		q_{\bm{\phi}}(\bm{z}|\bm{x}) = \prod_{i} q_{\bm{\phi}_i}(\bm{z}_i|\bm{z}_{i-1},\bm{x}_{i-k:i})\\
		\label{eq:HMVAE_decoder}
		p_{\bm{\theta}}(\bm{x}|\bm{z}) = \prod_{i} p_{\bm{\theta}_i}(\bm{x}_i|\bm{z}_{i-k:i})
	\end{align}
	with $\bm{x}_{i-k:i} = [\bm{x}_{i-k}^\mathrm{T},\ldots,\bm{x}_i^\mathrm{T}]^\mathrm{T}$, $\bm{z}_{i-k:i} = [\bm{z}_{i-k}^\mathrm{T},\ldots,\bm{z}_i^\mathrm{T}]^\mathrm{T}$. 
	The memory $k$ represents a trade-off between the expressiveness of the model and the extent to which the model is tailored to a particular task. In Fig.~\ref{fig:DVAE}~a) the corresponding probabilistic graph for $I=3$ and $k=1$ is shown. The dotted, dashed and solid arrows stand for the dependencies in $p_{\bm{\gamma}}(\bm{z})$, $q_{\bm{\phi}}(\bm{z}|\bm{x})$, and $p_{\bm{\theta}}(\bm{x}|\bm{z})$, respectively. If $k$ is set to 0, the independence assumptions equal those of a Kalman Filter and result in a highly sparse probabilistic graph. However, by increasing the parameter $k$, further dependencies are added and the sparsity level decreases. Since information about the distribution and temporal evolution of some scenario specific characteristics like doppler shifts can only be extracted from a sequence of channel observations, keeping the possibility of a nonzero $k$ is reasonable in the context of estimating wireless channels. By inserting \eqref{eq:DVAE_prior}, \eqref{eq:HMVAE_encoder} and \eqref{eq:HMVAE_decoder} into the definition of the \ac{ELBO} in~\eqref{eq:ELBO} , the new objective can be stated as
	\begin{equation}
		\label{eq:ELBO_DVAE}
		\mathcal{L}^{(\text{D})} = \sum_{n,i=1}^{N,I} \mathbb{E}_{q_{\bm{\phi}_{1:i}}}\left[\log \frac{p_{\bm{\theta}_i}(\bm{x}^{(n)}_i|\bm{z}_{i-k:i})p_{\bm{\gamma}_i}(\bm{z}_i|\bm{z}_{i-1})}{ q_{\bm{\phi}_i}(\bm{z}_i|\bm{z}_{i-1},\bm{x}^{(n)}_{i-k:i})}\right],
	\end{equation}
	where $q_{\bm{\phi}_{1:i}} = \prod_{i' = 1}^i q_{\bm{\phi}_{i'}}(\bm{z}_{i'}|\bm{z}_{i'-1},\bm{x}^{(n)}_{i'-k:i'})$. 
	Similiary to standard \acp{VAE}, the parameterized distributions $p_{\bm{\gamma}_i}(\bm{z}_i|\bm{z}_{i-1})$, $q_{\bm{\phi}_i}(\bm{z}_i|\bm{z}_{i-1},\bm{x}_{i-k:i})$ and $p_{\bm{\theta}_i}(\bm{x}_i|\bm{z}_{i-k:i})$ are modeled as Gaussians with means $\bm{\mu}_{\bm{\gamma}_i}(\bm{z}_{i-1})$, $\bm{\mu}_{\bm{\phi}_i}(\bm{z}_{i-1},\bm{x}_{i-k:i})$ and $\bm{\mu}_{\bm{\theta}_i}(\bm{z}_{i-k:i})$, and covariance matrices $\diag(\bm{\sigma}_{\bm{\gamma}}(\bm{z}_{i-1})^2)$, $\diag(\bm{\sigma}_{\bm{\phi}_i}(\bm{z}_{i-1},\bm{x}_{i-k:i})^2)$ and $\diag(\bm{\sigma}_{\bm{\theta}_i}(\bm{z}_{i-k:i})^2)$, respectively. Additionally, the expectation operations in \eqref{eq:ELBO_DVAE} are approximated by single sample Monte-Carlo estimations embedded in the encoder architecture. This is illustrated in Fig.~\ref{fig:DVAE}~b). The encoder is realized by $I$ ordered neural networks, where each one concatenates a \ac{CNN} with a \ac{FC} neural network and is followed by a sampling operation. The first encoder network outputs the mean $\bm{\mu}_{\bm{\phi_1}}(\bm{x}_{1})$ and variances $\bm{\sigma}_{\bm{\phi_1}}(\bm{x}_{1})$ of $q_{\bm{\phi}_1}(\bm{z}_1|\bm{x}_{1})$ from which a sample $\tilde{\bm{z}}_1$ is drawn from. This in turn is fed to the second encoder network computing $\bm{\mu}_{\bm{\phi}_2}(\tilde{\bm{z}}_1,\bm{x}_{1:2})$ and $\bm{\sigma}_{\bm{\phi}_2}(\tilde{\bm{z}}_1,\bm{x}_{1:2})$ such that a further sample $\tilde{\bm{z}}_2$ can be drawn from $q_{\bm{\phi}_{2}}(\bm{z}_{2}|\tilde{\bm{z}}_1,\bm{x}_{1:2})$. By repeating this procedure for all $I$ states we end up with a sample $\tilde{\bm{z}}$ drawn from $q_{\bm{\phi}}(\bm{z}|\bm{x})$ in \eqref{eq:HMVAE_encoder}. In this way, we can maximize the ELBO $\mathcal{L}^{(\text{D})}$ by a gradient based optimization algorithm over $\bm{\theta}$, $\bm{\phi}$ and $\bm{\gamma}$. The parameters $\bm{\theta}$ and $\bm{\gamma}$ are also implemented by $I$ distinct neural networks illustrated in Fig.~\ref{fig:DVAE}~c). A \ac{FC} neural network, called prior network, takes $\tilde{\bm{z}}_{i-1}$ as input, and outputs $\bm{\mu}_{\bm{\gamma}_i}(\tilde{\bm{z}}_{i-1})$ and $ \bm{\sigma}_{\bm{\gamma}_i}(\tilde{\bm{z}}_{i-1})$. On the other hand, another \ac{FC} neural network, called decoder network, outputs $\bm{\mu}_{\bm{\theta}_i}(\tilde{\bm{z}}_{i-k:i})$ and $\bm{\sigma}_{\bm{\theta}_i}(\tilde{\bm{z}}_{i-k:i})$ based on the input $\tilde{\bm{z}}_{i-k:i}$. The parameters $\log \bm{\sigma}_{\bm{\gamma}_i}$ and $\log \bm{\sigma}_{\bm{\phi}_i}$ are additionally bounded to improve the stability during training. In contrast to ordinary recurrent neural networks, no parameter sharing is utilized and there exist separate networks for each state $i$. The memory $k$, the latent dimension of $\bm{z}_i$, the widths and depths of the neural networks and the kernel size of the \ac{CNN} are determined via a network architecture search. The model leading to the largest \ac{ELBO} computed on an evaluation set is chosen.
	
	The \ac{kMMVAE} can be utilized in a similar way as the standard \ac{VAE} for \ac{ML} aided channel estimation. The dataset $\mathcal{X}$ consists of $I$ consecutive channel realizations along $N$ trajectories and the distributions $q_{\bm{\phi}_i}(\bm{z}_i|\bm{z}_{i-1},\cdot)$ are conditioned on noisy observations $\bm{y}_{i-k:i}$ of the channel realizations according to \eqref{aquisition_samples}. An estimation of the $i$-th channel realization $\bm{h}_i$ is obtained as follows. The first encoder network computes $\bm{\mu}_{\bm{\phi}_1}(\bm{y}_1)$ and forwards it to the next encoder network computing $\bm{\mu}_{\bm{\phi}_2}(\bm{\mu}_{\bm{\phi}_1},\bm{y}_{1:2})$. This in turn is used as input for the subsequent encoder network and is repeated up to the $i$-th time instance resulting in a sequence of means $\bm{\mu}_{\bm{\phi}_{1:i}} = [\bm{\mu}_{\bm{\phi}_1}^\mathrm{T},\ldots,\bm{\mu}_{\bm{\phi}_i}^\mathrm{T}]^\mathrm{T}$. Eventually, the $i$-th decoder network can compute $\bm{\mu}_{\bm{\theta}_i}(\bm{\mu}_{\bm{\phi}_{i-k:i}})$ and $\diag(\bm{\sigma}_{\bm{\theta}_i}(\bm{\mu}_{\bm{\phi}_{i-k:i}})^2)$. Based on these outputs, an approximated \ac{MMSE} channel estimator is obtained from \eqref{eq:LMMSE_MMSE}, where the expectation operation is dropped by replacing $\bm{\mu}_{\bm{h}|\bm{z}}$ and $\bm{C}_{\bm{h}|\bm{z}}$ with $\bm{\mu}_{\bm{\theta}_i}(\bm{\mu}_{\bm{\phi}_{i-k:i}})$ and $\diag(\bm{\sigma}_{\bm{\theta}_i}(\bm{\mu}_{\bm{\phi}_{i-k:i}})^2)$, respectively.
	
	\subsection{Related Channel Estimators}
	\label{sec:related_CE}
	We compare our proposed model to several other estimators. All methods are evaluated with respect to estimating the channel $\bm{h}_{\bar{i}}$ of one fixed time instance $\bar{i}$. As baseline for those which do not utilize \ac{ML}, we take the \ac{LS} estimator $\hat{\bm{h}}_{\bar{i}}^{(\text{LS})} = \bm{y}_{\bar{i}}$ and an \ac{LMMSE} estimator $\hat{\bm{h}}_{\bar{i}}^{(\text{sCov})}$ based on the sample covariance matrix $\hat{\bm{C}}_{\bar{i}}= (1/ N)\sum_{n=1}^{N} (\bm{h}_{\bar{i}}^{(n)}-\bar{\bm{h}}_{\bar{i}}^{(n)}) (\bm{h}_{\bar{i}}^{(n)}-\bar{\bm{h}}_{\bar{i}}^{(n)})^{\text{H}}$ with $\bar{\bm{h}}_{\bar{i}}^{(n)}$ being the sample mean. The dataset $\mathcal{X}_{\bar{i}}$ used in this case contains the channel realizations of the particular time instance $\bar{i}$ along all trajectories in $\mathcal{X}$, for which the channel estimation is evaluated. Additionally, we compare our model with the \ac{VAE} based estimator from \cite{VAEPaper}, which is explained in Section \ref{VAE_CE_instantenous_CSI} and for which the dataset $\mathcal{X}_{\bar{i}}$ is utilized as well. All these estimators solely rely on instantaneous \ac{CSI}. Therefore, we also consider a standard \ac{VAE}, called TSVAE, which takes a noisy observation of the whole trajectory as input and outputs the statistical characteristics of the channel at the time instance $\bar{i}$, which can be used to estimate $\bm{h}_{\bar{i}}$ in the same manner as it is described in Section \ref{VAE_CE_instantenous_CSI}.

	\section{Simulation Results}

	The generated trajectories in QuaDRiGa are simulated in a 3GPP 38.901 urban macro cell with mixed NLOS/LOS channels and 80\% indoor users. The number of snapshots per trajectory is set to 8 and the velocity for each user is drawn from a Rayleigh distribution with parameter $\sigma^2 = 4$. Motivated by the 5G standards, the center frequency is set to 2.1 GHz and the time interval $T$ between two successive snapshots is set to 0.5 ms. The number of antennas $R$ at the \ac{BS} is 32. 
	The data set is split into training, evaluation and test set with 100000, 10000 and 10000 trajectories, respectively. The channels along each trajectory are preprocessed separately by removing the average path gain over the snapshots as described in the QuaDRiGa documentation~\cite{QuadrigaDoc}. Moreover, the channels are normalized such that the average channel power of the 8-th snapshot $(1/ N)\sum_{n=1}^{N} \norm{\bm{h}_8^{(n)}}_2^2$ equals the number of \ac{BS} antennas. The noise variance is determined at a particular SNR for each trajectory individually by $\sigma_n^{(n)2} = \norm{\bm{h}_8^{(n)}}_2^2 / (R \cdot \text{SNR})$. New noisy observations $\bm{y}^{(n)}_i$ are generated for each epoch during the training. This is done by first drawing a value for the SNR in dB from a uniform distribution between -10 dB and 25 dB and then drawing a noise vector from $\mathcal{N}_{\mathbb{C}}(\bm{0},\sigma_n^{(n)2}\mathbf{I})$ for every snapshot along the trajectory. An adaptive learning rate is used which is initialized with $\mathrm{6 \cdot 10^{-5}}$ and once divided by a factor of 5 when the \ac{ELBO} remains constant within 25 epochs on the evaluation set. The \ac{ELBO} in \eqref{eq:ELBO_DVAE} is taken as objective which is additionally regularized by the method of free bits and maximized by the Adam optimizer~\cite{introVAE,adam}.
	
	The channel estimators are evaluated based on a snapshot-wise \ac{NMSE} defined as
	\begin{equation}
		\text{NMSE}_i = \frac{1}{N_t} \sum_{n=1}^{N_t} \frac{\norm{\bm{h}^{(n)}_i - \hat{\bm{h}}^{(n)}_i}_2^2}{\norm{\bm{h}^{(n)}_i}_2^2},
	\end{equation}
	where $N_t$ is the number of samples in the test set and $\hat{\bm{h}}^{(n)}_i$ is the estimate of $\bm{h}^{(n)}_i$.
	Fig.~\ref{fig:NMSE_over_SNR} shows the performance for an SNR range between -5 and 20 dB, where all methods are evaluated on the channel realizations of the 8-th snapshot. The data set $\mathcal{X}_8$ defined in Section \ref{sec:related_CE} is used for the estimators considering solely instantaneous \ac{CSI}. The \ac{kMMVAE} aided estimator clearly outperforms the other models for all SNR values. The model agnostic standard \ac{VAE} which takes the entire trajectory as input, called TSVAE, only performs slightly better than the standard \ac{VAE} considering only instantaneous \ac{CSI}. A reasonable explanation is that the TSVAE maps the trajectory on a standard normal distributed latent random vector $\bm{z}$ which cannot comprise temporal correlations efficiently. The \ac{LS} as well as the sample covariance estimator are generally worse than the \ac{ML} aided ones. This is expected since the sample covariance estimator is \ac{MSE} suboptimal for non-Gaussian distributed channels which usually holds. The \ac{LS} estimator does not incorporate any scenario specific information at all.
	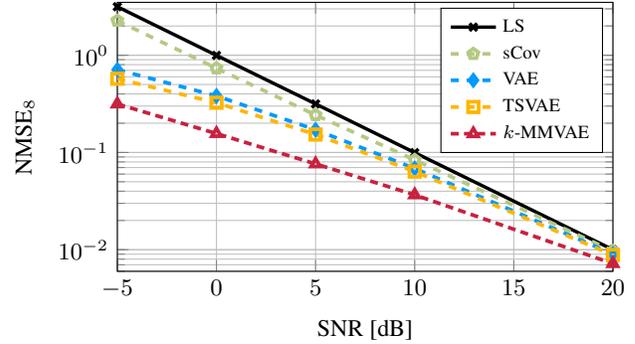
\begin{figure}[t]%
		\begin{tikzpicture}
	\definecolor{color1}{rgb}{0 0.4470 0.7410}
	\definecolor{color2}{rgb}{0.3010 0.7450 0.9330}
	\definecolor{color3}{rgb}{0.4940 0.1840 0.5560}
	\definecolor{color4}{rgb}{0.6350 0.0780 0.1840}
	\definecolor{color5}{rgb}{0.8500 0.3250 0.0980}
	\definecolor{color6}{rgb}{0.9290 0.6940 0.1250}
	\definecolor{color7}{rgb}{0 0.50 0}
	\definecolor{color8}{rgb}{0.4660 0.6740 0.1880}
	
	\definecolor{Yellow}{rgb}{1.00, 0.71, 0.00}
	\definecolor{Orange}{rgb}{1.00, 0.50, 0.00}
	\definecolor{Red}{rgb}{0.90, 0.20, 0.09}
	\definecolor{DarkRed}{rgb}{0.79, 0.13, 0.25}
	\definecolor{Blue}{rgb}{0.00, 0.60, 1.00}
	\definecolor{LightBlue}{rgb}{0.25, 0.75, 1.00}
	\definecolor{Green}{rgb}{0.57, 0.67, 0.42}
	\definecolor{LightGreen}{rgb}{0.71, 0.79, 0.51}

	\definecolorset{rgb}{TUMBeamer}{}{%
		Yellow,     1.00, 0.71, 0.00;%
		Orange,     1.00, 0.50, 0.00;%
		Red,        0.90, 0.20, 0.09;%
		DarkRed,    0.79, 0.13, 0.25;%
		Blue,       0.00, 0.60, 1.00;%
		LightBlue,  0.25, 0.75, 1.00;%
		Green,      0.57, 0.67, 0.42;%
		LightGreen, 0.71, 0.79, 0.51%
	}%
	
	\tikzset{VAE/.style={mark options={solid},color=Blue, line width=1.4pt, mark=diamond, mark size=2pt, dashed}}
	\tikzset{HMVAE/.style={mark options={solid},color=DarkRed, line width=1.4pt, mark=triangle, mark size=2pt, densely dashed}}
	\tikzset{TraVAE/.style={mark options={solid},color=Yellow, line width=1.4pt, mark=square, mark size=2pt, dashed}}
	\tikzset{LS/.style={mark options={solid},color=black, line width=1.3pt, mark=x, mark size=2pt, solid}}
	\tikzset{sCov/.style={mark options={solid},color=LightGreen, line width=1.4pt, mark=pentagon, mark size=2pt, dashed}}
	
	\newcommand{\legendVAE}         {\scriptsize VAE}
	\newcommand{\legendTSVAE}         {\scriptsize TSVAE}
	\newcommand{\legendLS}          {\scriptsize LS}
	\newcommand{\legendsCov}         {\scriptsize sCov}
	\newcommand{\legendHMVAE}     {\scriptsize $k$-MMVAE}
	\begin{axis}
		[width=0.95\columnwidth,
		height=0.6\columnwidth,
		xtick = {-5,0,5,10,15,20},
		xmin=-5,
		xmax=20,
		xlabel={SNR [dB]},
		ymode = log,
		ymin= 0.006,
		ymax=3.5,
		ylabel = {$\text{NMSE}_8$},
		grid = both,
		legend entries={
			\legendLS,
			\legendsCov,
			\legendVAE,
			\legendTSVAE,
			\legendHMVAE},
		legend cell align = {left},
		]
		\addplot[LS] table [x=SNR, y=LS, col sep=comma] {8snaps_NMSE_results.csv};
		\addplot[sCov] table [x=SNR, y=sCov, col sep=comma] {8snaps_NMSE_results.csv};
		\addplot[VAE] table [x=SNR, y=VAE, col sep=comma] {8snaps_NMSE_results.csv};
		\addplot[TraVAE] table [x=SNR, y=TraVAE, col sep=comma] {8snaps_NMSE_results.csv};
		\addplot[HMVAE] table [x=SNR, y=HMVAE, col sep=comma] {8snaps_NMSE_results.csv};
	\end{axis}
\end{tikzpicture}
		\caption{$\text{NMSE}_8$ over SNR of the different estimators. The \ac{kMMVAE} aided estimator is displayed in red.}
		\label{fig:NMSE_over_SNR}
	\end{figure}
	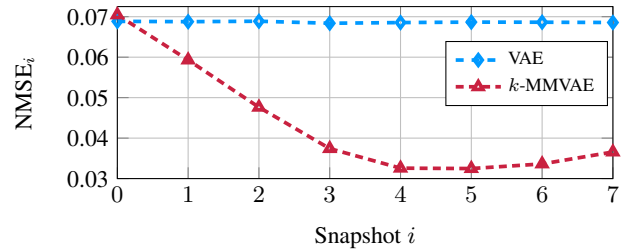
\begin{figure}[t]%
		\begin{tikzpicture}
	\definecolor{color1}{rgb}{0 0.4470 0.7410}
	\definecolor{color2}{rgb}{0.3010 0.7450 0.9330}
	\definecolor{color3}{rgb}{0.4940 0.1840 0.5560}
	\definecolor{color4}{rgb}{0.6350 0.0780 0.1840}
	\definecolor{color5}{rgb}{0.8500 0.3250 0.0980}
	\definecolor{color6}{rgb}{0.9290 0.6940 0.1250}
	\definecolor{color7}{rgb}{0 0.50 0}
	\definecolor{color8}{rgb}{0.4660 0.6740 0.1880}
	
	\definecolor{Yellow}{rgb}{1.00, 0.71, 0.00}
	\definecolor{Orange}{rgb}{1.00, 0.50, 0.00}
	\definecolor{Red}{rgb}{0.90, 0.20, 0.09}
	\definecolor{DarkRed}{rgb}{0.79, 0.13, 0.25}
	\definecolor{Blue}{rgb}{0.00, 0.60, 1.00}
	\definecolor{LightBlue}{rgb}{0.25, 0.75, 1.00}
	\definecolor{Green}{rgb}{0.57, 0.67, 0.42}
	\definecolor{LightGreen}{rgb}{0.71, 0.79, 0.51}
	
	\tikzset{VAE/.style={mark options={solid},color=Blue, line width=1.4pt, mark=diamond, mark size=2pt, dashed}}
	\tikzset{HMVAE/.style={mark options={solid},color=DarkRed, line width=1.4pt, mark=triangle, mark size=2pt, densely dashed}}
	
	\newcommand{\legendVAE}          {\scriptsize VAE}
	\newcommand{\legendHMVAE}     {\scriptsize $k$-MMVAE}
	\begin{axis}
		[width=0.95\columnwidth,
		height=0.45\columnwidth,
		xtick = {0,1,2,3,4,5,6,7},
		ytick = {0.03,0.04,0.05,0.06,0.07},
		yticklabels={0.03,0.04,0.05,0.06,0.07},
		scaled ticks = false,
		xmin=0,
		xmax=7,
		xlabel={Snapshot $i$},
		ymin= 0.03,
		ymax=0.0725,
		ylabel = {$\text{NMSE}_i$},
		grid = both,
		legend entries={
			\legendVAE,
			\legendHMVAE},
		legend style={at={(0.99,0.8)}},
		legend cell align = {left},
		]
		\addplot[VAE] table [x=SNAPSHOTS, y=VAE, col sep=comma] {8snaps_results.csv};
		\addplot[HMVAE] table [x=SNAPSHOTS, y=HMVAE, col sep=comma] {8snaps_results.csv};
	\end{axis}
\end{tikzpicture}
		\caption{NMSE over snapshots for 10 dB SNR. The memoryless \ac{VAE} aided estimator is displayed in blue. The \ac{kMMVAE} aided one is shown in red.}
		\label{fig:NMSE_over_snapshots}
		\vspace{-9pt}
	\end{figure}
	The \ac{kMMVAE} also provides the possibility to estimate all channels $\bm{h}_i$ along the trajectory which is shown in Fig.~\ref{fig:NMSE_over_snapshots}. The \ac{kMMVAE} as well as the VAE aided estimator are evaluated on every snapshot of the trajectory with a fixed SNR of 10 dB. Since the latter only considers instantaneous \ac{CSI} it performs the same in all cases. It can be seen how the \ac{kMMVAE} utilizes the knowledge of previous observations and improves the performance successively over the snapshots. For the last few time instances however, the \ac{NMSE} remains static and increases slightly. A possible explanation for this behaviour is, that the iterative sampling procedure explained in Section \ref{sec:HiddenMarkovVAE} leads to different importance levels of the learned parameters. More precisely, since the $i$-th drawn sample from $q_{\bm{\phi}_i}(\bm{z}_i|\tilde{\bm{z}}_{i-1},\bm{x}_{i-k:i}^{(n)})$ depends on the parameters $\bm{\phi}_{1:i-1}$, the encoder networks of the first few snapshots have more impact on which optimum is reached and are therefore adjusted more carefully than the encoder networks of the last snapshots. 
	
	The results indicate that \ac{ML} aided channel estimation can be improved significantly by incorporating the temporal evolution of the channel into the architecture. This, however, cannot be done in a straightforward and model agnostic fashion as it is done for the TSVAE estimator. Instead, a tailored method like the \ac{kMMVAE} aided estimator in which temporal characteristics can be learned efficiently, is beneficial.
	
	\section{Conclusion}
	In this work, we introduced a new DVAE architecture, called \ac{kMMVAE} and used it to extend the \ac{ML} aided channel estimator in~\cite{VAEPaper} to temporally correlated channels along trajectories. Our simulations showed that the model leads to notably better estimates compared to other \ac{ML} aided estimators. However, other applications like channel prediction for which the \ac{kMMVAE} could be considered have not yet been investigated. This topic together with a more detailed comparison to other \ac{DVAE} architectures and further classical estimators will be addressed in future work.
	
	\vfill\pagebreak
	
	\bibliographystyle{IEEEbib.bst}
	\bibliography{strings}
	
\end{document}